**Molecular-beam epitaxy of GaSb on 6°-offcut (001) Si using a GaAs nucleation layer**

M. Rio Calvo,[1] J-B. Rodriguez,[1] L. Cerutti,[1] M. Ramonda,[2] G. Patriarche,[3] and E. Tournié[1]

1. IES, Univ. Montpellier, CNRS, F-34000 Montpellier, France

2. CTM, Univ. Montpellier, F-34000 Montpellier, France

3. C2N, CNRS- Univ. Paris-Sud, Univ. Paris-Saclay, 10 Avenue Thomas Gobert, F-91120 Palaiseau, France

# Molecular-beam epitaxy of GaSb on 6°-offcut (001) Si using a GaAs nucleation layer

M. Rio Calvo,[1] J-B. Rodriguez,[1] L. Cerutti,[1] M. Ramonda,[2] G. Patriarche,[3] and E. Tournié[1]

1. IES, Univ. Montpellier, CNRS, F-34000 Montpellier, France
2. CTM, Univ. Montpellier, F-34000 Montpellier, France
3. C2N, CNRS- Univ. Paris-Sud, Univ. Paris-Saclay, 10 Avenue Thomas Gobert, F-91120 Palaiseau, France

**Abstract**

We studied and optimized the molecular beam epitaxy of GaSb layers on vicinal (001) Si substrates using a GaAs nucleation layer. An in-depth analysis of the different growth stages under optimized conditions revealed the formation of a high density of small GaAs islands forming a quasi-two-dimensional layer. GaSb then nucleated atop this layer as three-dimensional islands before turning to two-dimensional growth within a few nanometers. Moreover, reflexion high-energy electron diffraction revealed a fast relaxation of GaAs on Si and of GaSb on GaAs. The GaSb layer quality was better than that of similar layers grown on Si through AlSb nucleation layers.

1. **Introduction**

The combination of the silicon technology with the optical and electrical properties of the III-V semiconductors is under active consideration for developing novel optoelectronic devices, like photonic integrated circuits (PICs) or highspeed transistors. In this context, the integration of antimonides (GaSb, AlSb, InAs and their alloys and heterostructures) on silicon has recently attracted great attention since they offer short-, mid- and long-wavelength infrared (IR) emission and high electron/hole mobility [1]. Still, in spite of numerous studies, the direct epitaxial growth of III-V materials on silicon remains a challenge. The polar on non-polar growth, the large lattice mismatch and the thermal expansion coefficient difference are responsible for the formation of several types of crystallographic defects such as antiphase boundaries (APB), stacking faults, twins, threading dislocations (TD) or cracks, which drastically degrades the device performance. In addition, the III-V/Si interface energy always results in a Volmer-Weber growth mode, irrespective of the lattice mismatch [2,3]. Still, a considerable difference on islands shape and size has been observed with different III-V materials and growth conditions, which has been attributed to different interface energies [3] and kinetic-related processes such as the adatom diffusion, respectively [4].

Regarding the integration of antimonides, earlier studies have shown that GaSb nucleates on silicon as inhomogeneous large 3D islands [5] which results in low quality, high defect density GaSb layers after island coalescence. Machida *et al.* have recently demonstrated that the GaSb quality can be improved when it is directly grown on silicon if the conditions to nucleate a high density of small GaSb islands are fulfilled [5, 6]. On the other hand, inserting an AlSb nucleation layer at the interface improves the GaSb structural properties by the formation of a high density of small AlSb islands [7 - 11]. This approach has allowed us to demonstrate efficient quantum well [12] and quantum cascade [13] lasers grown on 6° offcut silicon substrates. The conclusion drawn from these works is that one should avoid the formation of large islands at the early stage of growth which can be achieved by kinetically restricting the system and the migration length of adatoms.

Following this idea, in this work we have investigated GaAs as a seed layer to grow GaSb layers on offcut Si substrates. The monolithic integration of GaAs compounds and Si technology has been extensively studied during the last years [14], and in particular many studies have focused on the growth initiation strategies [15,16]. Interestingly, there are three main differences between GaAs and AlSb NLs. On the one hand switching from the nucleation layer (NL) growth to the GaSb growth occurs by changing the group-V element (GaAs to GaSb) instead of the group-III element (AlSb to GaSb). The creation of Si-As bonds is expected to result in a different starting surface energy as compared to the AlSb NL [2]. On the other hand, while the lattice mismatch between AlSb and Si is around 13%, similar to the GaSb-to-Si mismatch, the mismatch between GaAs and Si is only around 4%. Therefore, in this approach the mismatch will be relieved in two steps, which might affect the growth as well.

2. **Experimental details**

The III-V materials have been directly grown on vicinal (001) Si substrates by Molecular Beam Epitaxy (MBE). $As_2$ and $Sb_2$ were provided by conventional cracker-cells. The Si substrates with a 6° offcut towards the [110] direction were used. They were first prepared *ex-situ* with cycles of HF dip and oxygen plasma, as described earlier [17]. Subsequently, the samples were flash annealed at 800°C inside the MBE without any intentional impinging flux, in order to remove residual impurities from the surface. Substrate temperatures higher than 400°C were measured with an infrared pyrometer, while lower ones were inferred from the thermocouple. The growth were monitored through *in-situ* reflection high energy electron diffraction (RHEED). The epitaxial samples were characterized by atomic force microscopy (AFM) and high-resolution X-Ray diffraction (HRXRD). The surface morphology was measured in tapping mode using a Nanoman AFM microscope, controlled by a Nanoscope V electronics from Bruker Instruments, using a cantilever with a nominal tip radius between 2 and 5 nm (Nanosensors NCL tip). The roughness - namely the root mean square (rms) of the distribution of height - and power spectral density function (PSDF) were extracted from the AFM topographic images using the Gwyddion software [18]. HRXRD measurements were carried out using a PANalytical X'Pert MRD equipped with a four bounce (002) Ge monochromator and an X-Ray tube providing the Cu k$\alpha_1$ radiation. The omega-scans were recorded using a PIXcel linear detector in the open detector configuration, which correspond to a 2.5° aperture angle. Scanning transmission electron microscopy (STEM) observations were made on a Titan Themis 200 microscope (FEI/ Thermo Fischer Scientific) equipped with a geometric aberration corrector on the probe. This microscope is also equipped with the "Super-X" systems for EDX analysis with a detection angle of 0.9 steradian. The observations are made at 200 kV with a probe current of about 70 pA and a half-angle of convergence of 24 mrad. HAADF-STEM images are acquired with a camera length of 110 mm (inner/outer collection angles are respectively 69 and 200 mrad). The thin cross-sections were prepared by FIB, the cuts are made along the [110] zone axis, which is the axis of disorientation.

The growth were performed in two steps: First a GaAs NL and a 50 nm-thick GaSb layer were grown at the same temperature. Next, the substrate temperature was changed, with the sample surface left under Sb flux, and the structure was completed by a 450 nm-thick GaSb layer grown at 500 °C, the usual growth temperature of GaSb. The gallium flux was kept constant for both the GaAs and GaSb layers, and set to get an equivalent GaSb growth rate of 0.3 ML/sec. When not otherwise specified the V/III flux ratio was around 2 for all layers.

3. Results and discussion

*3.1 Influence of the nucleation layer growth parameters on the GaSb quality*

In this section, we report on the influence of the growth parameters used for the GaAs nucleation layer on the structural properties of the GaSb layer grown atop. The V/III flux ratio, the substrate growth temperature, the shutter sequence and the GaAs layer thickness were varied.

We have first studied the influence on the GaSb quality of the As/Ga ratio during the growth of GaAs. A set of samples was grown with As/Ga ratio in the 1.3 to 4 range and characterized using HRXRD. The sample structure as well as the evolution of the FWHM of the rocking curves are presented in Fig. 1.

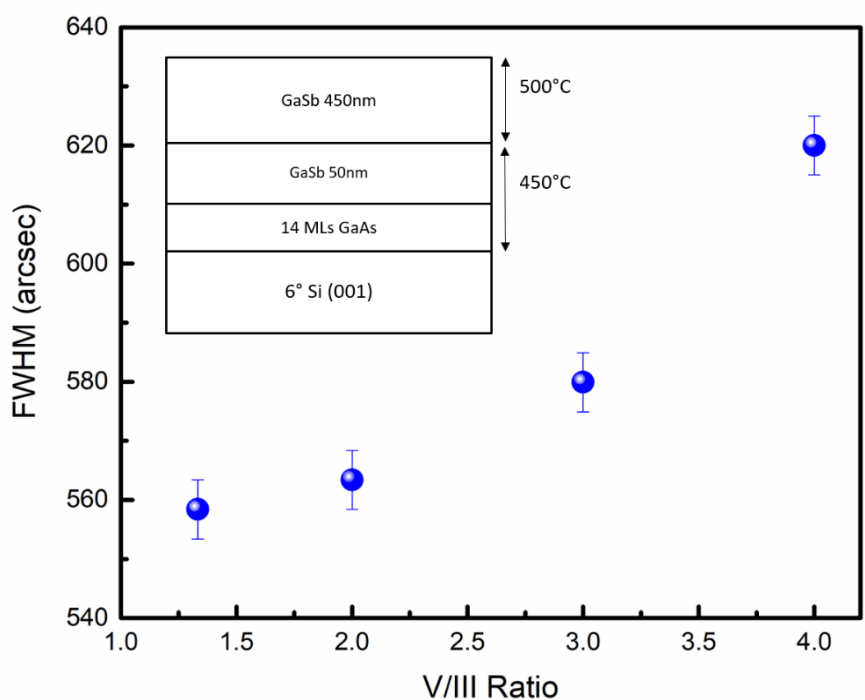

**Fig 1.** Variation of the FWHM of the 004 GaSb peaks and the AFM roughness with the V/III ratio of the GaAs nucleation layer. Inset, scheme of the growth with the temperatures.

The results clearly show that an As/Ga flux ratio in the 1 to 2 range gives the best GaSb quality, with a FHWM around 560 arcsec. For larger As fluxes, the FWHM rapidly increases. A low V/III ratio during the GaAs growth therefore seems necessary, which we ascribe to two possible explanations: On the one hand, the substrate temperature at this stage of the growth sequence is low as compared to the common temperatures used for the growth of GaAs (~600°C), and a too large excess of As may result in a severe degradation of the material quality. On the other hand, when the growth of GaSb is initiated, the residual As sitting at the GaAs surface and the As background in the MBE reactor is likely to be incorporated into the GaSb layer. In fact, it was found that in the case of GaSb grown on GaAs substrate, the best quality is achieved when the As-to-Sb swap is performed at high temperature in order to replace residual As by Sb and therefore avoid any incorporation of As into GaSb [19,20]. In an attempt to further reduce the amount of residual excess As at the GaAs/GaSb interface, a 20" Sb soak of the GaAs surface was introduced before starting the GaSb growth, which only resulted in a minor improvement of the FWHM (540 arcsec compared to 560 arcsec without Sb soak).

Next, the effect of the substrate temperature during the growth of the GaAs NL and the first 50 nm-thick GaSb layer was investigated by growing a set of samples at various NL temperatures in the 370 to 450°C range with an As/Ga flux ratio of 1.3. Fig. 2 shows that the best GaSb quality is achieved when GaAs and the low temperature GaSb are grown near 400 °C. A rather small variation is observed at lower temperature whereas it degrades significantly at high temperatures (nearly 50 arcsec more). The AFM images measured on NLs grown at 450°C (Fig. 3) reveal an island morphology with an average island diameter around 20 nm and a rms of 2.9 nm due to the presence of tall islands with an average size between 5 and 6 nm, and a density of around $5 \times 10^8$ cm$^{-2}$.

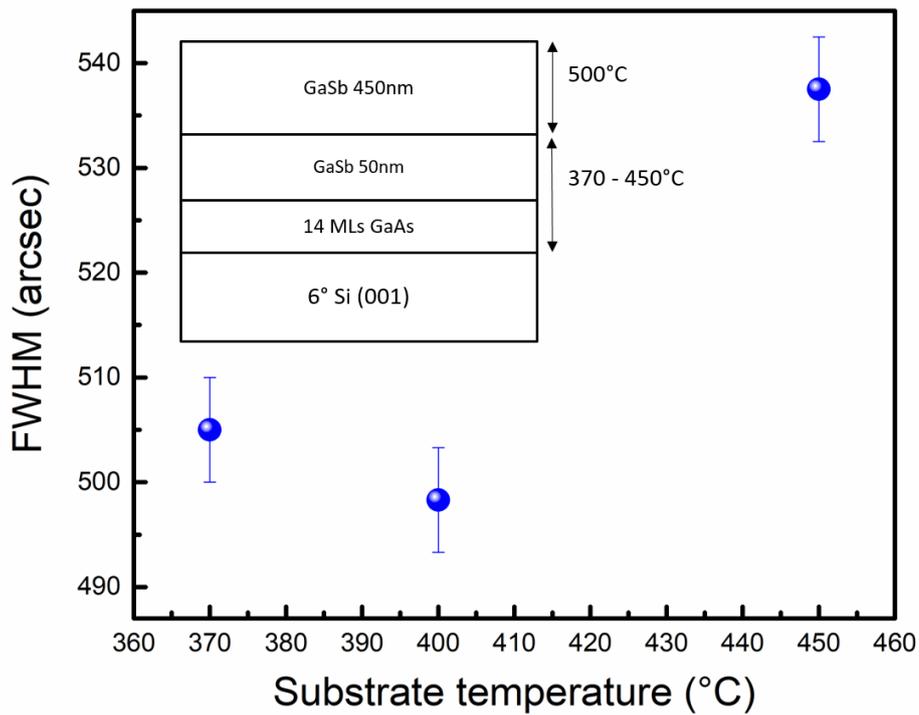

**Fig 2**. Variation of the FWHM of the 004 GaSb peaks with the GaAs substrate temperature during the nucleation layer.

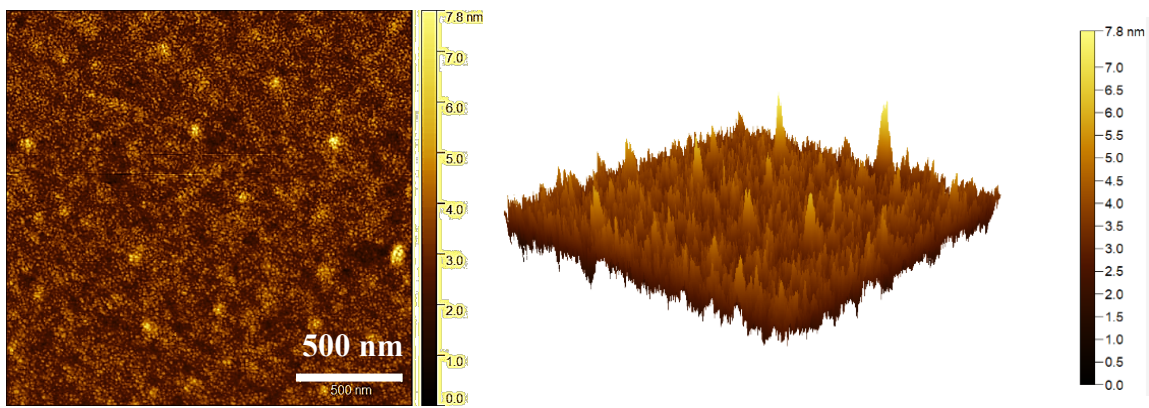

**Fig 3**. Planar (left) and 3D (right) views of AFM (2 µm x 2 µm) images taken from 14 MLs GaAs grown on Si at 450°C.

Finally, two sets of samples were grown to investigate the impact of the GaAs NL thickness at two growth temperatures (400 and 450°C respectively) and two As/Ga flux ratio (1.3 and 4 respectively). The rocking curve FWHMs and AFM rms roughnesses obtained on 20 x 20 µm² images are displayed in Fig. 4 (a) and (b), respectively.

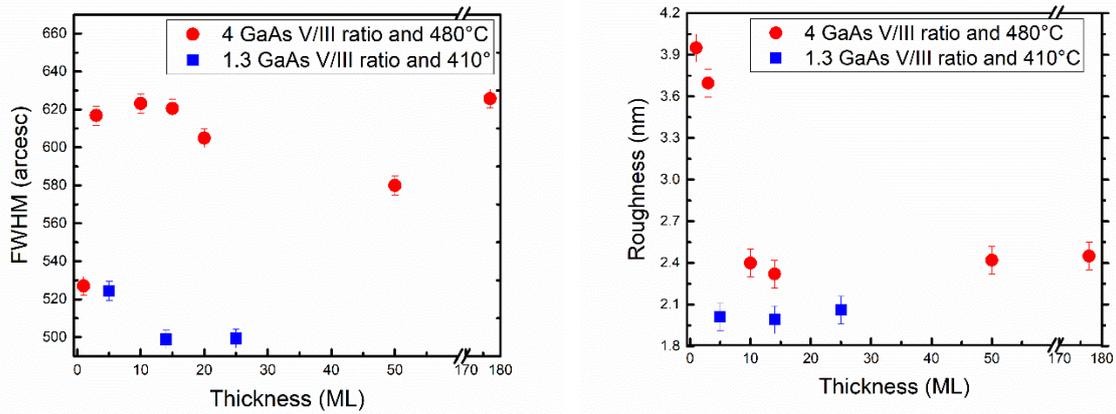

**Fig 4**. Variation of the FWHM of the 004 GaSb peaks and the AFM roughness with the GaAs nominal thickness for different substrates heater temperatures and GaAs V/III ratios.

As expected from the experiments described above, the set of samples where the NL has been grown at low temperature using a low As/Ga flux ratio demonstrate better FWHM and rms roughness than the set grown at high temperature and large As/Ga ratio. However, the difference between the two sets is much more pronounced on the FWHM (nearly 100 arcsec), which could be ascribed to a better arrangement of the misfit dislocation array at the GaAs/Si or GaSb/GaAs interfaces, or to a faster dislocation recombination rate within the layers.

From these graphs, an optimal NL thickness can be derived for each set of samples: about 14 MLs at low temperature and low As/Ga ratio, and about 50 MLs in the other case. It is interesting to note here a similarity with a work we previously published with AlSb NLs [9]. Indeed, in this work, we found that the optimal NL thickness is critically dependent on the substrate temperature, which we attributed to the onset of the coalescence of the AlSb islands. This optimal thickness increases with the temperature, as it seems to be the case here, too. However, the most interesting feature in the present case is the fact that for GaAs NL thicker than 5 MLs, the variation of the roughness and the FWHM are very weak compared to the AlSb case. In fact, in the thickness range investigated here (5 to 25 MLs (5 to 180 MLs) in the low (high) temperature / low (high) V/III ratio (respectively)), the FWHM values only varies by 20 to 40 arcsec and the rms roughness by less than 0.2 nm. This relative insensitivity to the GaAs NL thickness clearly contrasts with AlSb for which far much larger variations were observed. In both cases however, the substrate temperature and the V/III ratio must be carefully adjusted in order to improve the material quality. It is also noteworthy that, as far as HR-XRD and AFM are concerned, the material quality obtained using a GaAs NL is better than when an AlSb NL is used. Indeed, optimal growth conditions using the later resulted in a FWHM of 580 arcsec and a rms roughness of 4 nm [10], to be compared with 490 arcsec and about 2 nm respectively with the GaAs NL. Finally, in plane polar figures (IPPF) measured on the samples presented here revealed that the twined volume in the GaSb layers is below the detection sensitivity limit of our setup (not shown here for the sake of conciseness). This again contrasts with AlSb NL which were shown to generate twins when grown at low temperature [21].

### *3.2 Post growth annealing*

Finally, the impact of post growth annealing on the structural properties of the layers has been investigated. Complete structures with GaAs NL thicknesses of 14 MLs grown at two different substrate temperatures, namely 400 and 450°C, were annealed and characterized. Right after the growth, HRXRD measurements were carried out and the samples were reloaded in the MBE system to perform annealing during 1 h at 550°C under an Sb flux. This procedure was repeated twice. Figure 5 shows the evolution of the 004 GaSb peak FWHM with the annealing duration. Our best results previously obtained using an AlSb NL [9] were also added to the graph for the sake of comparison. After 2 h annealing, the samples comprising a GaAs NL grown at low temperature exhibit an FWHM as low as 420 arcsec, which is significantly better than the value obtained with an AlSb NL (490 arcsec) [9]. This figure confirms that a GaAs NL allows the growth of better GaSb quality than an AlSb NL, as well as that low temperature should be used for the NL.

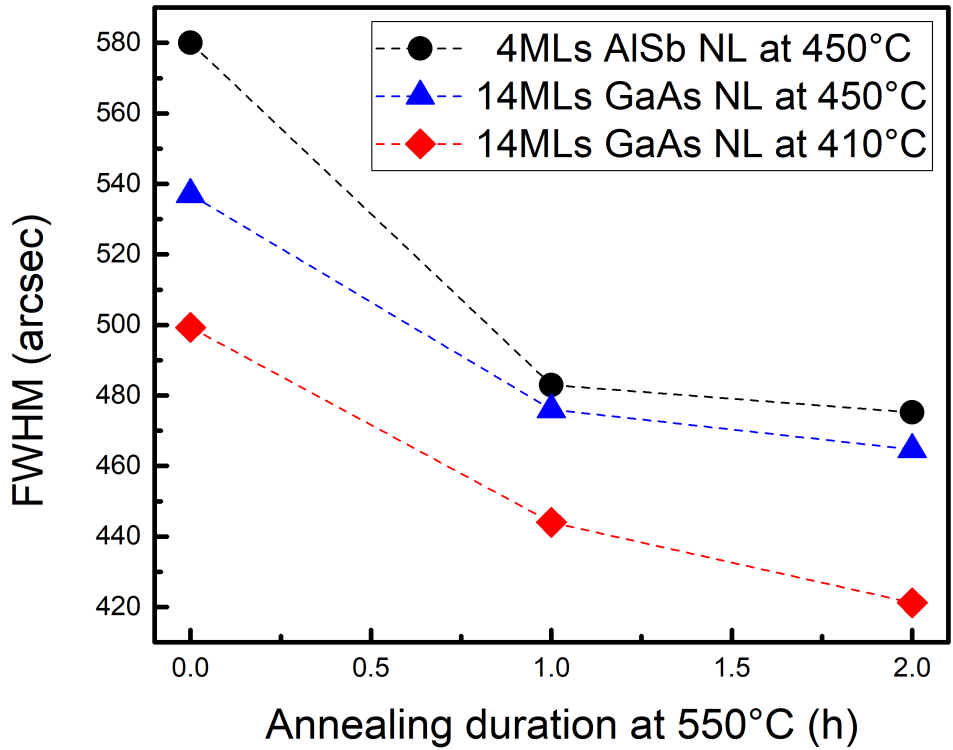

Fig 5. Variation of the FWHM with the annealing duration for different GaAs substrate temperatures and comparison with AlSb nucleation layer thickness.

## 4   Microstructure of the GaSb on GaAs-on-Si samples.

In this section, we study in detail the structural properties of a sample grown using the optimized parameters derived from the experiments described above, namely: a 14 MLs thick GaAs NL grown using a V/III ratio of 1.3 at a substrate temperature of 400°C, a 50 nm thick layer of GaSb grown at the same temperature, finally covered by a 450 nm thick GaSb layer grown at 500°C.

Notably, no change in the (2x1) silicon reconstruction was observed prior to the growth [15] while a spotty RHEED pattern was observed during the growth of GaAs, as shown in the inset of Fig.6 (a), indicating the formation of islands, as expected for the VW growth mode of III-Vs on Si. The AFM picture taken on a NL sample reveals an overall homogeneous surface with a low rms roughness of 0.65 nm (Fig. 6a). We will come back to this rather surprising observation later on. After deposition of 1.5 ML GaSb (Fig. 6b), the spots on the RHEED pattern appeared brighter and less elongated. The AFM image taken on the surface of a sample corresponding to this stage of the growth shows the same uniform coverage of the Si surface due to the NL. However, it also reveals brighter areas with an average size of 50 nm and an average separation of 260 nm which we ascribe to GaSb islands atop the NL. A (1x3) reconstruction characteristic for GaSb growth appeared after a few MLs only and was superimposed to the spotty RHEED pattern (Fig.6c). The AFM image of the surface of a sample with 50 MLs GaSb demonstrates a rougher surface due to the emergence of big elongated GaSb islands. This behavior is comparable to what is observed when growing GaSb on GaAs substrates [2]. The RHEED pattern then progressively became stricky, attesting the transition to the growth of a two-dimensional, smoother layer. After the full growth sequence is completed, the AFM shows a flat surface with a rms roughness as low as 2 nm (Fig.6d), which can be ascribed to the presence of the emerging TDs visible on this picture as dark spots, with a density in the 3 to 5 x$10^9$ cm$^{-2}$ range.

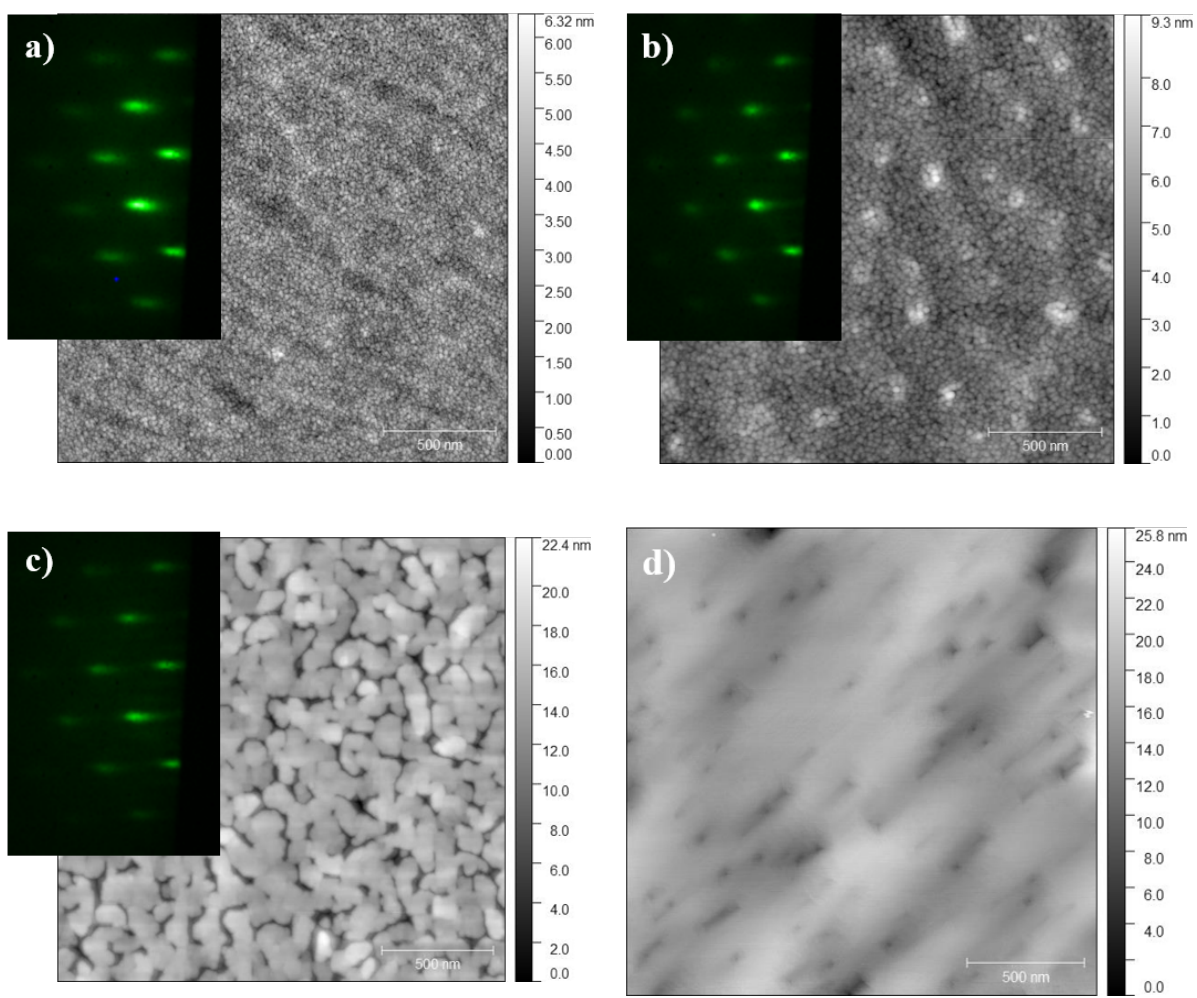

**Fig 6.** In-situ RHEED pattern and (5 µm x 5 µm) AFM topographic images. a) 14 MLs GaAs (rms: 0.65 nm). b) 1.5 ML GaSb – 14 MLs GaAs (rms: 1.26 nm). c) 50 MLs GaSb – 14 MLs GaAs (rms: 2.6 nm). d) 450 nm GaSb - 50 nm GaSb – 14 MLs GaAs (rms: 2 nm).

We combined AFM and TEM measurements to investigate the actual morphology of the NL. Indeed, higher magnification AFM image of the NL (Fig.7a) shows uniform and densely packed islands forming a quasi-continuous layer, which explains the observation of a continuous layer seen in Fig. 6 a. The power spectral density function indicates an average island diameter of 20 nm, and a density of $\sim 10^{11}\ cm^{-2}$, in agreement with previously reported results [13]. This uniform coverage of the Si surface by a large density of small GaAs islands is further confirmed by HAADF- and EDX- STEM images. The images presented in Fig.7b, c and d show such a measurement carried out on a sample comprising both the NL and the 500 nm thick GaSb layer. The GaAs islands distribution can clearly be identified on the elementary EDX chemical mapping of arsenic (Kα line) displayed on Fig.7d. The islands form a quasi-continuous 2D layer, with a typical width of 15 to 20 MLs and a rather flat top surface. Interestingly, from the comparison between these results and the morphology measured by AFM directly after the GaAs was grown, it can be deduced that the GaSb overgrowth does not significantly changes the GaAs nucleation layer morphology.

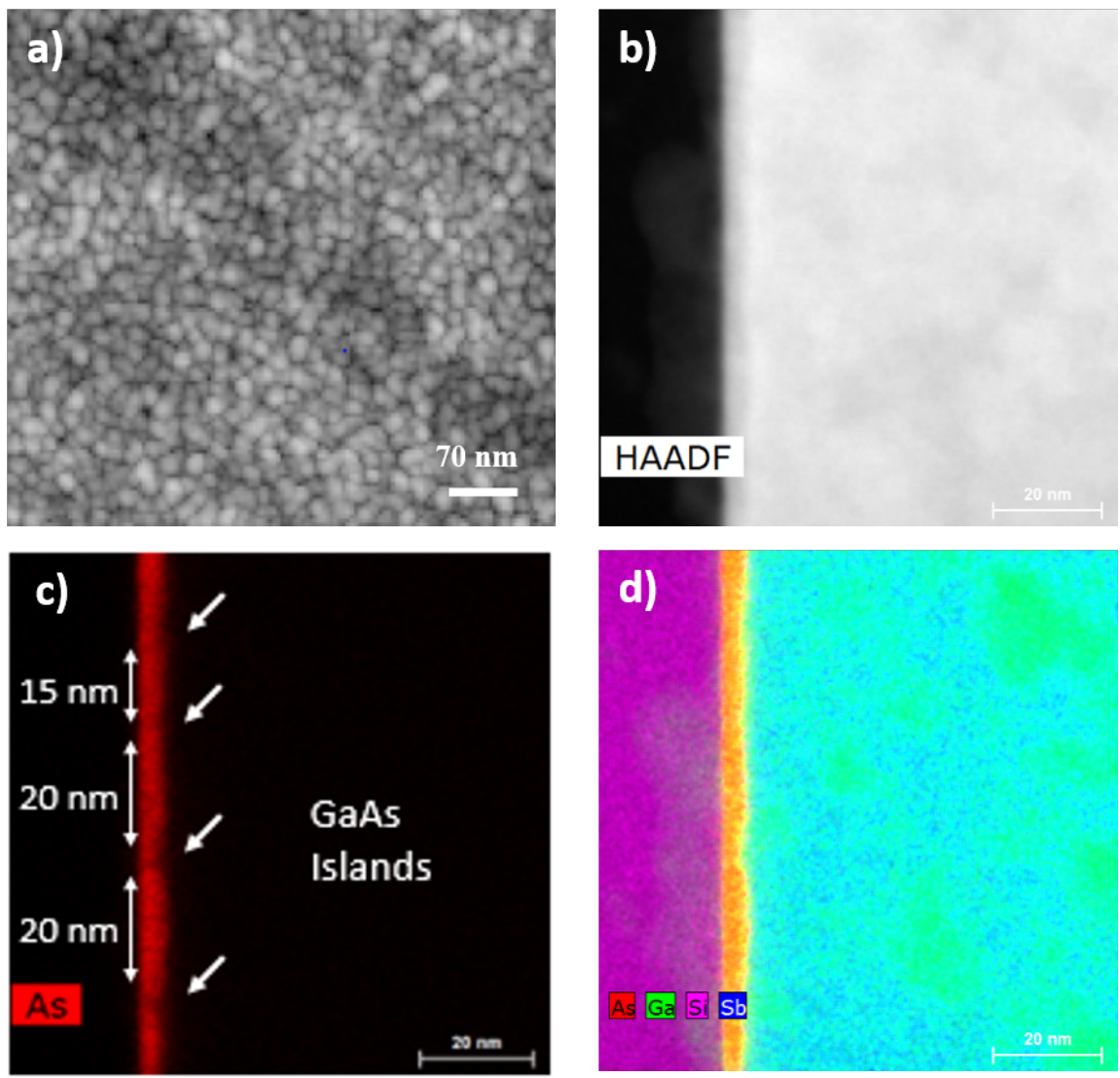

**Fig 7** a) (2 μm x 2 μm) AFM topographic images. b) HAADF–STEM image c) EDX elemental mapping with the As map d) the superimposed chemical mapping of the elements As, Ga, Si and Sb .

Next, we have recorded the evolution of the RHEED spots spacing during the growth of the GaAs NL and part of the GaSb layer grown at low-temperature. This measurement allows describing the evolution of the lattice parameter at the growth front [20], and the result obtained is shown on Fig.8. A rapid evolution of the lattice parameter toward that of GaAs was observed during the NL deposition, full relaxation of the 4% mismatch being achieved after ~ 8 MLs. Still, a careful inspection of Fig. 8 reveals a two-step relaxation of the GaAs layer. This is ascribed to an initial elastic relaxation stage through island formation, followed by the generation of dislocations to fully relieve the strain, as reported by others [23]. In contrast, the 8% mismatch between GaSb and GaAs was relieved very rapidly in a single step, and full relaxation was achieved after around 4 MLs (Fig. 8). This relaxation is similar to what is observed when growing GaSb on GaAs substrates and can thus be ascribed to the generation of a network of misfit dislocations at the interface [22].

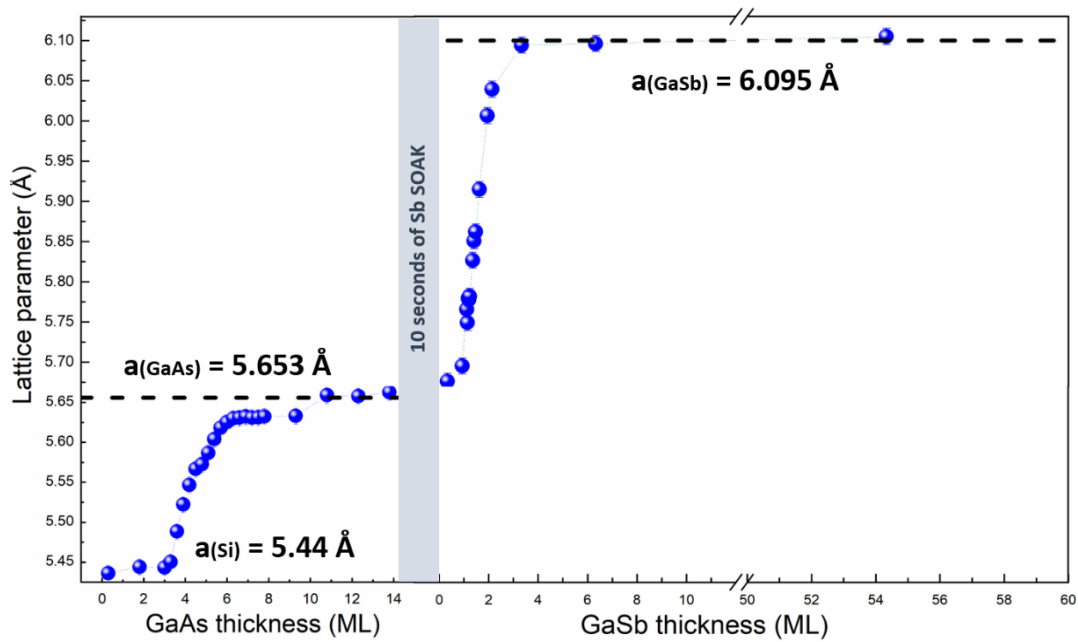

**Fig 8.** 50 nm GaSb/14 MLs GaAs/6° Si (001) grown at 400°C. RHEED measure of in plan lattice parameter in the vertical direction to the Si surface. Lattice parameter variation with the thickness, for 14 MLs of GaAs and 53 MLs of GaSb on the left and right side, respectively.

For a better comprehension of the relaxation process and the dislocation network, the maps of the in-plane strain fields have been obtained by Geometric Phase Analysis from the high resolution HAADF-STEM images. They reveal the presence of dislocations at the two interfaces Si/GaAs as well as GaAs/GaSb. The two misfit dislocations networks are very irregular. The cores of the dislocations are often poorly defined because of the low coherence length of the dislocation along the direction [110]. Some dislocations are also dissociated (the stacking faults between the two partial dislocations appear as yellow segments on the $\varepsilon_{xx}$ deformation map). The low growth temperatures used for the growth can explain the poor organization of the two misfit dislocations networks.

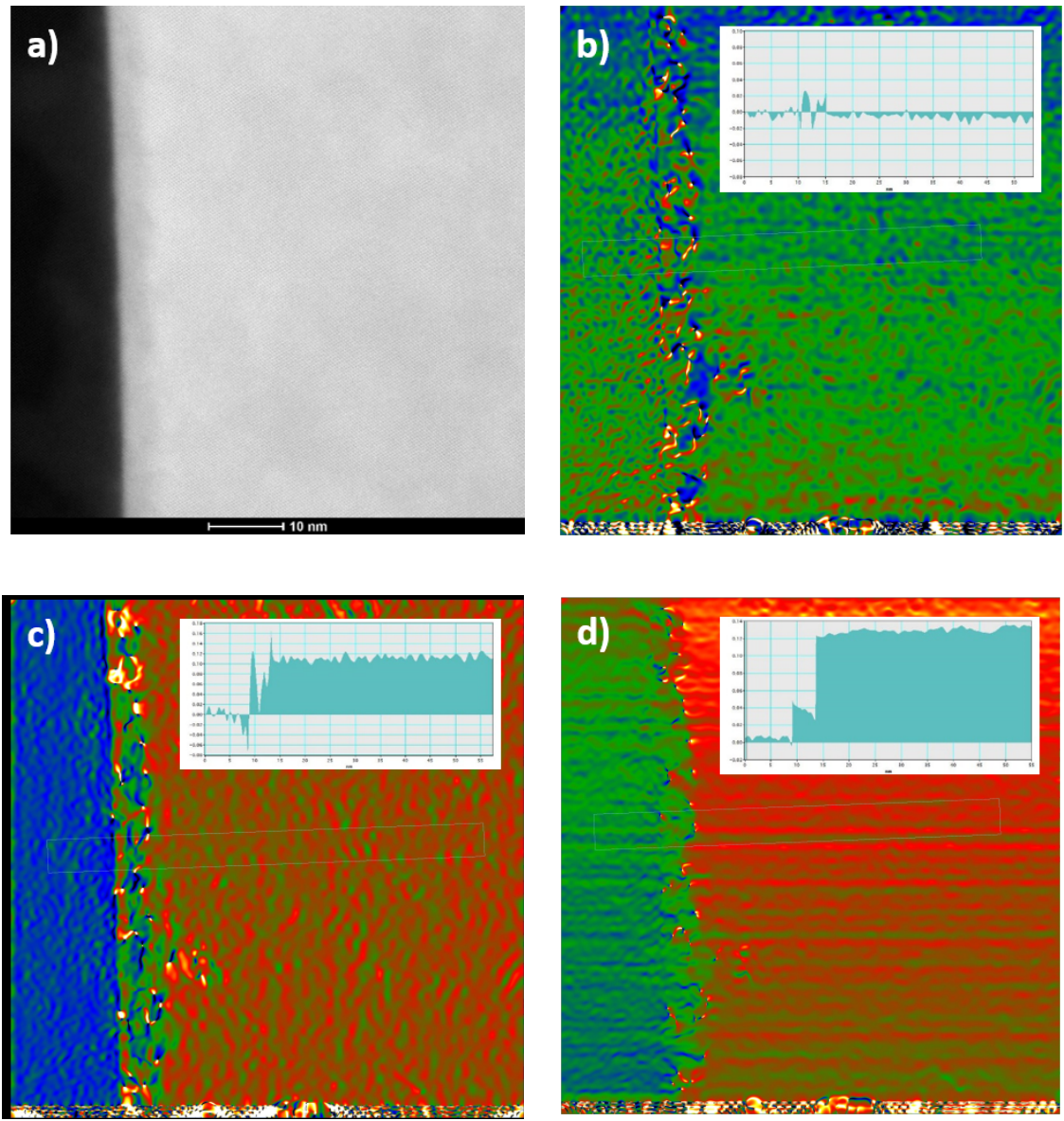

**Fig 9.** 500 nm of GaSb on 14MLs GaAs on (001) Si 6°. Atomic resolution HAADF-STEM image (a) quantitative strain fields measurements obtained by Geometrical Phase Analysis treatment. Maps of the in-plan strain fields: $\varepsilon_{xy}$ (b), $\varepsilon_{xx}$ (c) and $\varepsilon_{yy}$ (d). The x axis is aligned in the [001]

direction (the growth direction) and the y axis in the [1-10] direction. The thin foil is prepared in the [1-10] zone axis.

## 5 Conclusions

In summary, we have investigated the epitaxial growth of GaSb on 6°-off silicon substrates using GaAs as a nucleation layer. RHEED, X-Ray diffraction techniques, AFM and TEM measurements were performed in order to characterize the different growth stages and the samples quality. This study demonstrated that using a low As/Ga ratio as well as a low temperature during GaAs NL growth significantly improved the quality of the subsequent GaSb layer, whereas the NL thickness was not as critical as when an AlSb NL was used. The results of an in-depth analysis of the different growth stages, and in particular the NL deposition, revealed the formation of a high density of small GaAs islands (~ $10^{11}\ cm^{-2}$) forming a quasi 2D layer. The *in-situ* lattice relaxation measurements performed by RHEED indicate a rapid strain relaxation at both interfaces, GaAs on the Si substrate and GaSb on the GaAs NL. Finally, the quality of the samples obtained using a GaAs NL (FWHM of 420 arcsec) was significantly better than the value obtained with an AlSb NL (490 arcsec) which is promising in view of the monolithic integration of GaSb-based devices on Si.

**Acknowledgment**

Part of this work was sponsored by the French program on "Investments fort he Future" (EquipEX EXTRA, ANR-11-EQPX-0016) and the H2020 program of the European Union (REDFINCH, GA 780240).